\documentclass[runningheads]{llncs}
\usepackage{graphicx}
\usepackage{xcolor}
\usepackage{siunitx}
\usepackage{bm}
\usepackage{amsmath}
\usepackage{amssymb}
\usepackage{booktabs}
\usepackage[linesnumbered,ruled]{algorithm2e}
\usepackage{mathtools}
\usepackage{url}

\begin{document}
\title{Domain-agnostic segmentation of thalamic nuclei from joint structural and diffusion MRI}
\titlerunning{Thalamic segmentation networks}
%
\author{Henry~F.~J.~Tregidgo\inst{1} \and
Sonja~Soskic\inst{1} \and
Mark~D.~Olchanyi\inst{2,3} \and
Juri~Althonayan\inst{1} \and
Benjamin~Billot\inst{1,4} \and
Chiara~Maffei\inst{5} \and
Polina~Golland\inst{4} \and
Anastasia~Yendiki\inst{5} \and
Daniel~C.~Alexander\inst{1} \and
Martina~Bocchetta\inst{6,7} \and
Jonathan~D.~Rohrer\inst{7} \and
Juan~Eugenio~Iglesias\inst{1,4,5} } 
\authorrunning{H.F.J. Tregidgo et al.}
%
%
\institute{Centre for Medical Image Computing, UCL, London, UK\\
\email{h.tregidgo@ucl.ac.uk} \and
Neurostatistics Research Laboratory, MIT, Cambridge, USA\and
Center for Neurotechnology and Neurorecovery, MGH, Boston, USA\and
Computer Science and Artificial Intelligence Laboratory, MIT, Cambridge, USA\and
Martinos Center for Biomedical Imaging, Harvard Medical School, Boston, USA \and
Centre for Cognitive and Clinical Neuroscience, Brunel University, London,~UK \and
Dementia Research Centre, UCL, London, UK}
\maketitle              
%

\begin{abstract}
The human thalamus is a highly connected subcortical grey-matter structure within the brain. 
It comprises dozens of nuclei with different function and connectivity, which are affected differently by disease.
For this reason, there is growing interest in studying the thalamic nuclei \emph{in vivo} with MRI. 
Tools are available to segment the thalamus from 1 mm T1 scans, but the contrast of the lateral and internal boundaries is too faint to produce reliable segmentations. 
Some tools have attempted to incorporate information from diffusion MRI in the segmentation to refine these boundaries, but do not generalise well across diffusion MRI acquisitions.  
Here we present the first CNN that can segment thalamic nuclei from T1 and diffusion data of any resolution without retraining or fine tuning. 
Our method builds on a public histological atlas of the thalamic nuclei and silver standard segmentations on high-quality diffusion data obtained with a recent Bayesian adaptive segmentation tool. 
We combine these with an approximate degradation model for fast domain randomisation during training. 
Our CNN produces a segmentation at 0.7 mm isotropic resolution, irrespective of the resolution of the input. 
Moreover, it uses a parsimonious model of the diffusion signal at each voxel (fractional anisotropy and principal eigenvector) that is compatible with virtually any set of directions and b-values, including huge amounts of legacy data.
We show results of our proposed method on three heterogeneous datasets acquired on dozens of different scanners. An  implementation of the method is  publicly available at \url{https://freesurfer.net/fswiki/ThalamicNucleiDTI}.
\end{abstract}


\section{Introduction}

The human thalamus is a brain region with connections to the whole cortex~\cite{Schmahmann:2003ty,Behrens:2003aa}. 
It comprises dozens of nuclei that are involved in diverse functions like cognition, memory, sensory, motor, consciousness, language, and  others~\cite{Sherman:2001tz,Schmahmann:2003ty,Fama:2015wo}. 
Crucially, these nuclei are differently affected by disease, e.g., the  intralaminar nuclei are involved in Parkinson’s~\cite{Henderson:2000vj},  the anterior nuclei in Alzheimer's~\cite{Braak:1991tj,Braak:1991uy}, or the pulvinar in frontotemporal dementia~\cite{Vatsavayai:2016wf}. 
Such differentiation has sparked interest in studying the thalamic nuclei \emph{in vivo} with MRI.
This requires automated segmentation methods at the subregion level, as opposed to the whole thalamus provided by neuromaging packages like FreeSurfer~\cite{Fischl:2002tl} or FSL~\cite{Patenaude:2011vy}, or by convolutional neural networks (CNNs) like DeepNAT~\cite{Wachinger:2018tk} or SynthSeg~\cite{billot2023synthseg}. 

Different approaches have been used to segment thalamic nuclei. 
Some methods have attempted to register manually labelled histology to MRI~\cite{Krauth:2010vk,Jakab:2012ta,Sadikot:2011to}, but accuracy is limited by the difficulty in registering two modalities with such different contrasts, resolutions, and artifacts. 
Diffusion MRI (dMRI) has been used to spatially cluster voxels into subregions, based on similarity in diffusion signal~\cite{Mang:2012uk,Battistella:2017vn,Semedo:2018ui} or connectivity to cortical regions~\cite{Behrens:2003aa,Johansen-Berg:2005aa}. 
Clustering based on functional MRI connectivity has also been explored~\cite{Zhang:2008aa}. 
However, such clusters are not guaranteed to correspond to anatomically defined nuclei.

Other methods have relied on specialised MRI sequences to highlight the anatomical boundaries of the thalamus, typically at 7T~\cite{Tourdias:2014tk,Liu:2020tu} or with advanced dMRI acquisitions~\cite{Basile:2021te}. 
A widespread method within this category is ``THOMAS'', a labelled dataset of 7T white-matter-nulled scans that has been used to segment the thalamic nuclei with multi-atlas segmentation~\cite{Su:2019aa} and CNNs~\cite{Umapathy:2021vd}. 
The main disadvantage of THOMAS is requiring such advanced acquisitions at test time, which precludes the analysis of data acquired at sites without the required resources or expertise, as well as of legacy data acquired with standard sequences. 

One approach that seamlessly supports training and test data of different modalities is Bayesian segmentation, which combines a probabilistic atlas (derived from one modality) with a likelihood model to compute adaptive segmentations on other modalities using Bayesian inference~\cite{Ashburner:2005aa}. 
A probabilistic atlas of thalamic nuclei built from 3D reconstructed histology is available on FreeSurfer, along with a companion Bayesian segmentation method to segment the nuclei from 1~\si{mm} isotropic T1-weighted scans~\cite{Iglesias:2018aa}. 
An improved version of this method that incorporates dMRI  into the likelihood model for more accurate segmentation has recently been released~\cite{Tregidgo:2022aa}, but it inherits the well-known problems of Bayesian segmentation with partial voluming (PV)~\cite{van2003unifying}.
While this tool works well with high-resolution dMRI data (like  the Human Connectome Project, or HCP~\cite{Sotiropoulos:2013aa}), the lack of PV modelling is detrimental at resolutions much lower than $\sim$1~\si{mm} isotropic; this is the case of virtually every legacy dataset, and many modern datasets that use  lower resolutions (e.g., to keep acquisition time short, or for consistency with older timepoints in longitudinal studies, like ADNI~\cite{Jack-Jr.:2008wl}). 

Finally, there are also supervised discriminative methods that label the thalamus from dMRI. 
An early approach by Stough and colleagues~\cite{stough2014automatic} used a random forest to segment the thalamus into six groups of nuclei. 
As features, they used local measures like fractional anisotropy (\emph{FA}) and the principal eigenvector (\emph{V1}), and connectivity with cortical regions. 
A recent approach~\cite{ewert2021learning} segmented the whole thalamus (and other brain regions) using a CNN with six input channels -- one per unique element of the diffusion tensor image (\emph{DTI}) at each voxel. 
While these supervised approaches can provide excellent performance on the training domain, they falter on datasets  with different resolution. 

Here, we present the first CNN that can segment the thalamic nuclei from dMRI and T1 data out of the box, without retraining or finetuning. 
We use domain randomisation to model  resolution  during training, which enables the CNN to produce super-resolved 0.7~\si{mm} isotropic segmentations, independently of the resolution of the input images. 
Aggressive data augmentation is used to ensure robustness against variations in contrast, shape and artifacts. 
Finally, our CNN uses a  parsimonious representation of the dMRI data (FA+V1), which makes our publicly available tool compatible with virtually every dMRI dataset.


\begin{figure}[t]
    \centering
    \includegraphics[width=0.86\textwidth]{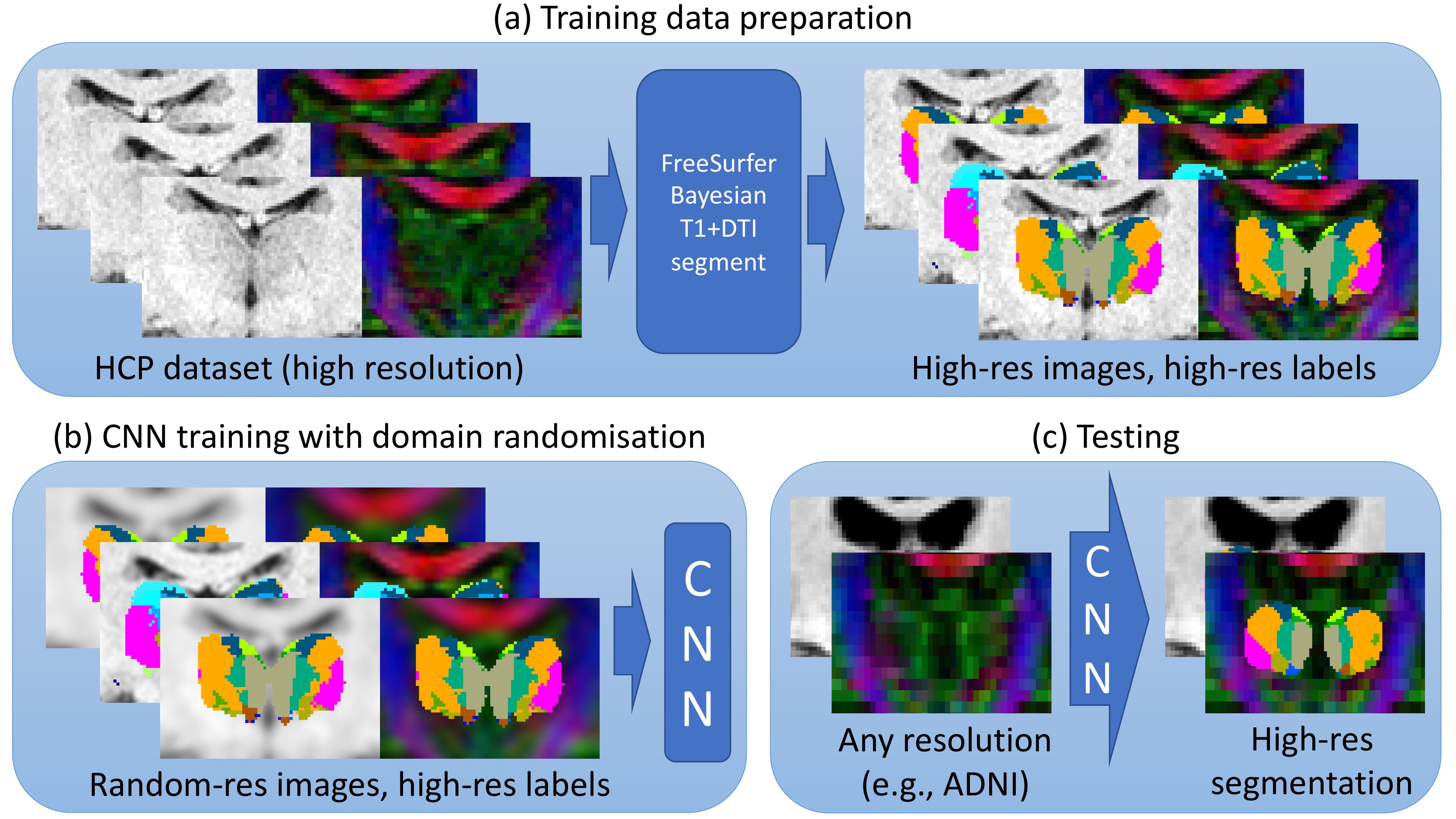}
    \caption{Overview of the proposed method. (a)~Generation of labelled training data. (b)~Training with domain randomisation. (c)~Testing. Images are in coronal view.}
    \label{fig:overview}
\end{figure}

\section{Methods}

A summary of our method is shown in Figure~\ref{fig:overview}. 
We use the joint T1/DTI Bayesian method in FreeSurfer to segment the thalamic nuclei from a large number of modern, high-quality scans. 
These segmentations are used as silver standard to train a CNN, thus circumventing the need for manual segmentations. Our approach uses a hybrid domain randomisation and augmentation strategy that enables the network to generalise to virtually any diffusion dataset. 

\subsection{Training dataset, preprocessing, and data representation}
\label{sub:trainingset}

To make the CNN compatible with legacy datasets, we choose a simple representation based on the FA and V1 of the DTI fit at each voxel. 
DTI only requires 7 measurements and is thus compatible with even the oldest datasets. 
As in many DTI visualisation tools, we combine the FA and V1  into a single 3$\times$1 red-green-blue vector at every voxel. 
This RGB vector has brightness proportional to the FA, and its colour encodes the direction of V1 as shown in Figure~\ref{fig:overview}a.

To obtain accurate training segmentations from the Bayesian method in FreeSurfer~\cite{Tregidgo:2022aa} we require a high-resolution dataset with reduced PV artifacts. 
We choose the HCP, which includes
0.7~\si{mm} isotropic T1 and 1.25~\si{mm} isotropic dMRI with 90 directions and three b-values (1000, 2000, and 3000~\si{s/mm^2}).
We use the HCP to generate our targets and then generate training images at a wide spectrum of resolutions by increasing the voxel size with a degradation model.
We consider two RGB images per subject, derived from DTI fits of the b=1000 and b=2000 shells, respectively. For each of the two DTI fits, the Bayesian method yields three different sets of segmentations  (corresponding to three available likelihood models). All six segmentations are defined on the .7~\si{mm} grid (Figure~\ref{fig:overview}a, right), and comprise 23 thalamic nuclei per hemisphere (46 total)~\cite{Tregidgo:2022aa}.

\subsection{Domain randomisation and data augmentation}
\label{sub:augmentation}
We employ domain randomisation and aggressive data augmentation for both our T1 and diffusion data in order to model: \textit{(i)}~the degradation in quality from HCP to more standard acquisition protocols, and \textit{(ii)}~the variability in appearance due to differences in acquisitions and scanners at test time. 

\vspace{2mm}
\noindent\textbf{Domain randomisation for resolution:} 
at the crux of our method is the domain randomisation of input resolutions. 
At every iteration, we randomly sample the voxel dimensions for the T1 and DTI (independently) in two steps. 
First, we sample a ``coarse'' scalar voxel size from a uniform distribution between  $1$ and $3~\si{mm}$. 
Then, we sample the voxel side length in each direction from a normal distribution centred on this ``coarse'' mean with $\sigma=0.2~\si{mm}.$ 

Next, we resample the T1 and RGB channels to the sampled resolution. 
For the T1, we use a publicly available PV-aware degradation model~\cite{Billot:2020wf}, which accounts for variability in slice thickness and slice spacing. 
For the RGB, one should theoretically downsample the original diffusion-weighted images, and recompute the DTI at the target resolution. 
However, the exact characteristics of the blurring depend on the set of directions and b-values, which will not be the same for the training and test datasets. Moreover, recomputing the DTI is too slow for on-the-fly augmentation. 
Instead, we apply the degradation model to the RGB image directly, which can be done very efficiently. 
This is only an approximation to the actual degradation, but in practice, the domain randomisation strategy minimises the effects of the domain gap created by the approximation.

\vspace{2mm}
\noindent\textbf{Data augmentations:} we also apply a number of geometric and intensity augmentations, some standard, and some specific to our dMRI representation.

\noindent\textit{- Global geometric augmentation:} 
we use random uniform scaling between 0.85 and 1.15, and random uniform rotations about each axis between -15 and 15 degrees. Rotations are applied to the images and also used to reorient the V1 vectors.

\noindent\textit{- Local geometric augmentation:} 
we deform the scans with a piecewise linear deformation field, obtained by linear interpolation on a $5\times5\times5\times3$ grid.  V1 vectors are reoriented with the PPD method (``preservation of principle direction''~\cite{Alexander:2001aa}). 

\noindent\textit{- Local orientation augmentation:} 
we further generate a smooth grid of random rotations between $[-15^{\circ},15^{\circ}]$ around each principal axis using piecewise linear interpolation on a $5\times5\times5\times3$ grid. 
These are applied to each vector in V1 to simulate noise and variations in principle direction

\noindent\textit{- DTI ``speckles'':} 
To account for infeasible FA and V1 voxels generated by potentially unconstrained DTI fitting, we select random voxels in the low resolution images (with probability $p=1\times10^{-4}$), randomise their RGB values, and renormalise them so that their effective FA is between 0.5 and 1.

\noindent\textit{- Noise, brightness, contrast, and gamma:} 
we apply random Gaussian noise to both the T1 and FA; randomly stretch the contrast and modify the brightness of the T1; and  apply a random gamma transform to the  T1 and FA volumes. 

\vspace{2mm}
We note that these augmentations are applied to the downsampled images. After that, the augmented images are upscaled back to 0.7 mm isotropic resolution. 
This ensures that all the channels (including the target segmentations) are defined on the same grid, independently of the intrinsic resolution of the inputs (Figure~\ref{fig:overview}b). 
At test time, this enables us to produce .7 mm segmentations for scans of any resolution (prior upscaling to the .7 mm grid).

\subsection{Loss}
\label{sub:loss}
We build on the standard soft dice loss~\cite{Milletari:2016vv}, which is widely used in segmentation with CNNs. 
Since some labels in the histological atlas are very small, we implemented a grouped soft dice by combining nuclei into 10 larger labels. 
We borrowed these groupings from ~\cite{Tregidgo:2022aa}. 
We then combined this Dice with the average Dice of the individual nuclei and the Dice of the whole thalamus into the following composite loss:
\begin{small}
\begin{equation}
\mathcal{L} = - \sum_lSDC(X_l,Y_l) - \sum_gSDC\left(\{X_l\}_{G_g},\{Y_l\}_{G_g}\right) - SDC\left(\{X_l\}_{l\neq0},\{Y_l\}_{l\neq0}\right),
\label{eq:loss}
\end{equation}
\end{small}
where, $X_l=\{x^l_i\}$ and $Y_l=\{y^l_i\}$ are the predicted and ground truth probability maps for label $l\in [0,\ldots,L]$; $G_g$ is the set of label indices in nuclear group $g\in[1,\ldots,10]$, label $l=0$ corresponds to the background and $SDC$ is the soft Dice coefficient: 
$ SDC(X,Y) = (2\times\sum_ix_iy_i) / (\sum_ix_i^2+\sum_iy_i^2). $

\subsection{Architecture and implementation details}
Our CNN is a 3D U-net~\cite{ronneberger2015u,cciccek2016_3d} with 5 levels (2 layers each),  3$\times$3$\times$3 kernels and ELU activations~\cite{clevert2015fast}. 
The first level has 24 features, and every level has twice as many features as the previous one. 
The last layer has a softmax activation. 
The loss in Equation~\ref{eq:loss} was optimised for 200,000 iterations with Adam~\cite{kingma2014adam}. 
A random crop of size 128$\times$128$\times$128 voxels (guaranteed to contain the thalami) was used at every iteration. 
The T1 scans are normalised by scaling the median white matter intensity to 0.75 and clipping at [0, 1]. 
The DTI volumes are upsampled to the 0.7~\si{mm} space of the T1s (using the log domain~\cite{Arsigny:2006vj}) prior to the RGB computation. 
To generate a training target we combine all six segmentation candidates (three likelihood models times two shells) in a two step process: \textit{(i)}~averaging the one-hot encodings of each segmentation,  and\textit{(ii)}~coarsely segmenting into 10 label groups and renormalising the soft target.

For validation purposes, we used the Bayesian segmentations of 50 withheld HCP subjects and 14 withheld ADNI subjects. 
Even though the Bayesian segmentation of ADNI data is not reliable enough to be used as ground truth for evaluation  (due to the PV problems described in the Introduction), it is still informative for validation purposes -- particularly when combined with HCP data. 
The final model for each network is chosen based on the validation loss averaged between the HCP and ADNI validation sets.


\section{Experiments \& Results}
\begin{figure}[t]
    \centering
    \includegraphics[width=0.79\textwidth]{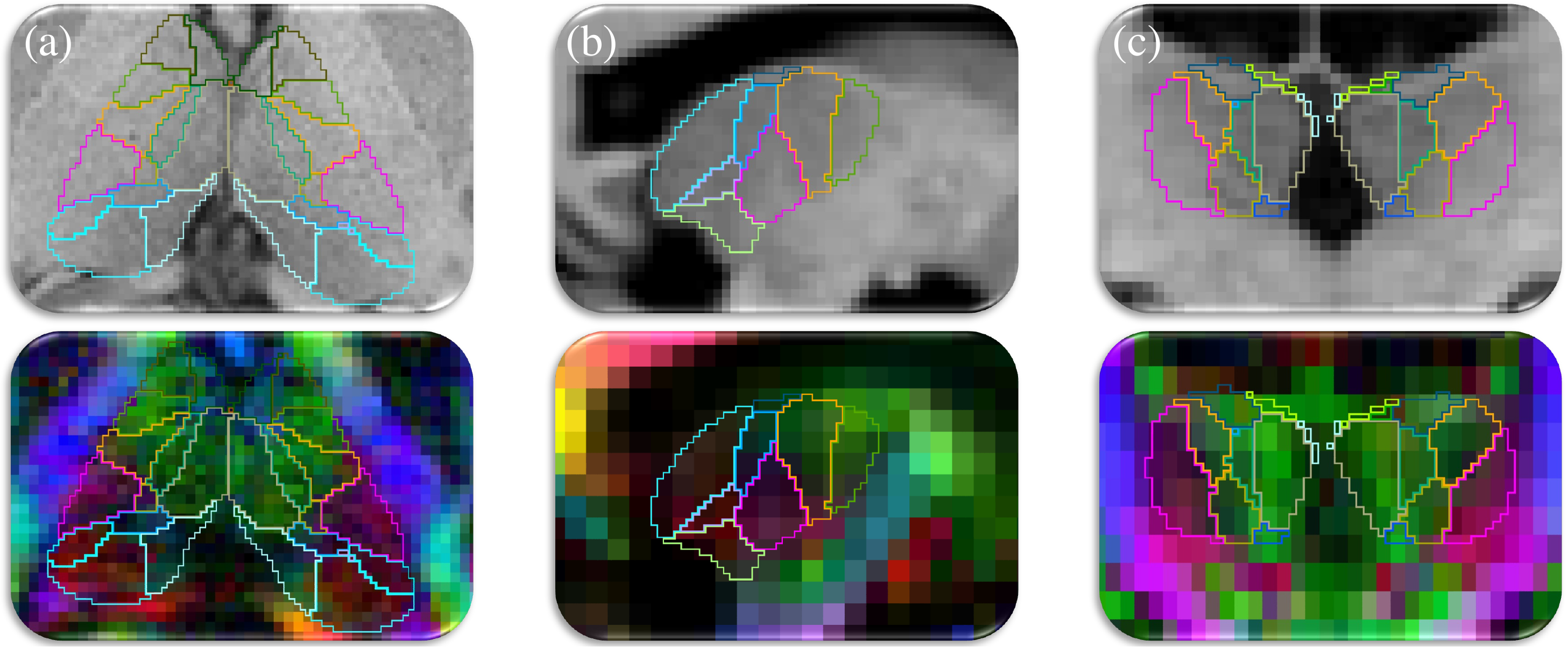}
    \caption{Example segmentations from: (a)~HCP (axial);  (b)~LOCAL  (sagittal); and (c)~ADNI (coronal). Top: T1. Bottom: RGB encoding. The CNN produces accurate segmentations at high isotropic resolution despite the heterogeneous acquisitions.}
    \label{fig:examples}
\end{figure}

\subsection{MRI data}
We trained our CNN on 600 subjects from the HCP dataset, as explained in Section~\ref{sub:trainingset}. For evaluation, we used three different datasets:

\vspace{1mm}
\noindent\textbf{HCP:}
10 randomly selected subjects (not overlapping with the training data), with manual segmentations of 10 groups of labels (the same as in Section~\ref{sub:loss}).

\vspace{1mm}
\noindent\textbf{LOCAL:} 
21 healthy subjects (9 males, ages 53-80), each with a $1.1~\si{mm}$ isotropic T1 and a test-retest pair of 2.5~\si{mm} isotropic dMRI (64 directions, b=1,000).

\vspace{1mm}
\noindent\textbf{ADNI:} 
90 subjects from the ADNI, 45 with with Alzheimer's disease (AD) and 45  healthy controls (73.8$\pm$7.7 years; 44 females), each with a T1  (1.2$\times$1$\times$1~\si{mm}, sagittal) and dMRI (1.35$\times$1.35$\times$2.7~\si{mm}, axial, 41 directions, b=1,000).

\subsection{Competing methods and ablations}
To the best of our knowledge, the only available thalamic segmentation tool for T1/dMRI that can segment data of arbitrary resolution is the Freesurfer Bayesian tool. We therefore compare our method with this algorithm, along with seven ablations of model options: using only the Dice loss; three variations on the way of merging the three candidate Bayesian segmentations into a training target (average one-hot; majority voting; and selecting a segmentation at random); and three ablations on the augmentation (omitting the ``speckle'' DTI voxels, the random V1 rotations, and the piecewise linear deformation).

\subsection{Results}

Qualitative results are shown in Figure~\ref{fig:examples}, which displays sample segmentations for the three datasets. Our CNN successfully segments all scans at 0.7~\si{mm} resolution, despite the different voxel sizes of the inputs. Quantitative results are presented below for three experimental setups, one with each dataset.

\vspace{1mm}
\noindent\textbf{Direct evaluation with manual ground truth using HCP:} 
We first evaluated all competing methods and ablations on the 10 manually labelled subjects. Table~1 (left columns) shows mean Dice scores at different levels of granularity. 
Thanks to the ground truth aggregation, domain randomisation and aggressive augmentation, most of the CNNs produce higher accuracy than the Bayesian method at every level of detail -- despite having been trained on automated segmentations from the Bayesian tool. The only ablation with noticeable lower performance is the one using the Dice of only the fine histological labels (i.e., no Dice of groupings), which highlights the importance of our composite Dice loss.

\begin{table}[t]
    \label{tab:TRT_GT}
    \centering
    \includegraphics[width=0.86\textwidth]{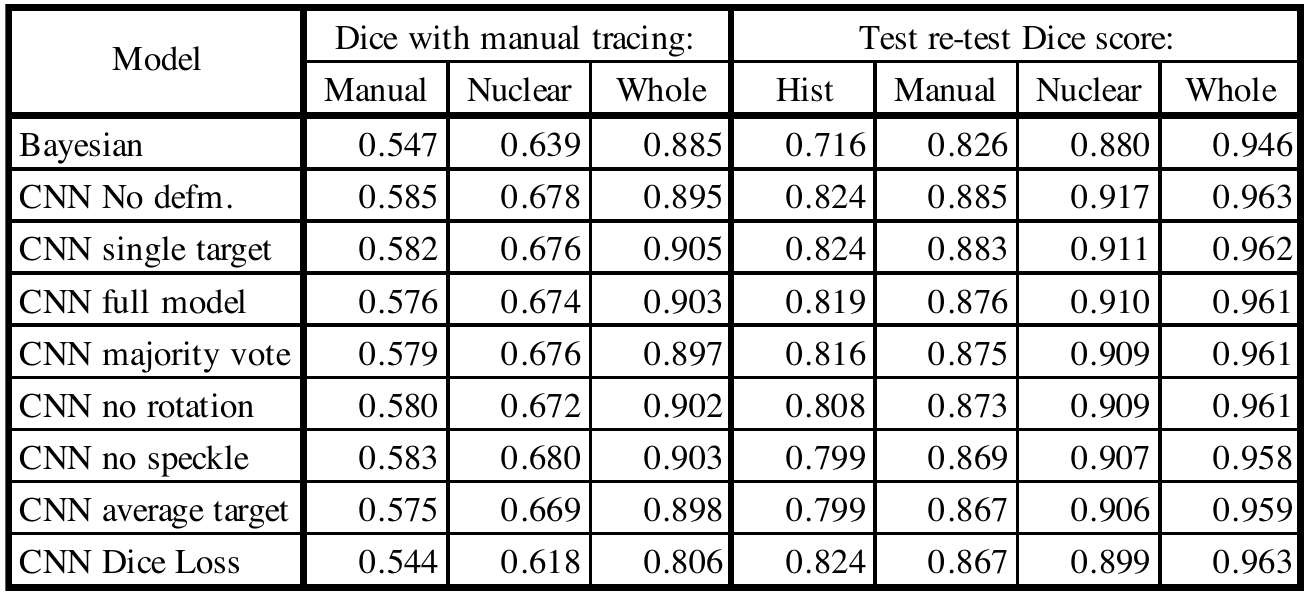}
    \vspace{2mm}
    \caption{Mean Dice for ground truth comparison (left columns) and test re-test (right columns). Dice is shown for labels grouped into: histological labels (``hist'', 23 labels), manual protocol (``manual'', 10 labels), nuclear groups~\cite{Tregidgo:2022aa} (``nuclear'', 5 labels), and  whole thalamus. CNNs are sorted in descending order of average Dice across columns.}
\end{table}

\vspace{2mm}
\noindent\textbf{Test-retest using LOCAL:} 
Table~1 (right columns) shows Dice scores between the segmentations of the two available sets of images for the LOCAL dataset, for different levels of granularity. All the networks are more stable than the Bayesian method, with considerably higher test-retest dice scores. 

\vspace{2mm}
\noindent\textit{Best-performing CNN:} Analysing the results from Table~1 as a whole, the CNN with the highest mean Dice across the board is the one without local geometric augmentation. We hypothesise that this is because the benefit of this augmentation is negligible due to the large number of training cases (600), and thus does not compensate for the loss of performance due to the approximations that are required to augment on the fly.

\begin{figure}[t]
    \centering
    \includegraphics[width=0.91\textwidth]{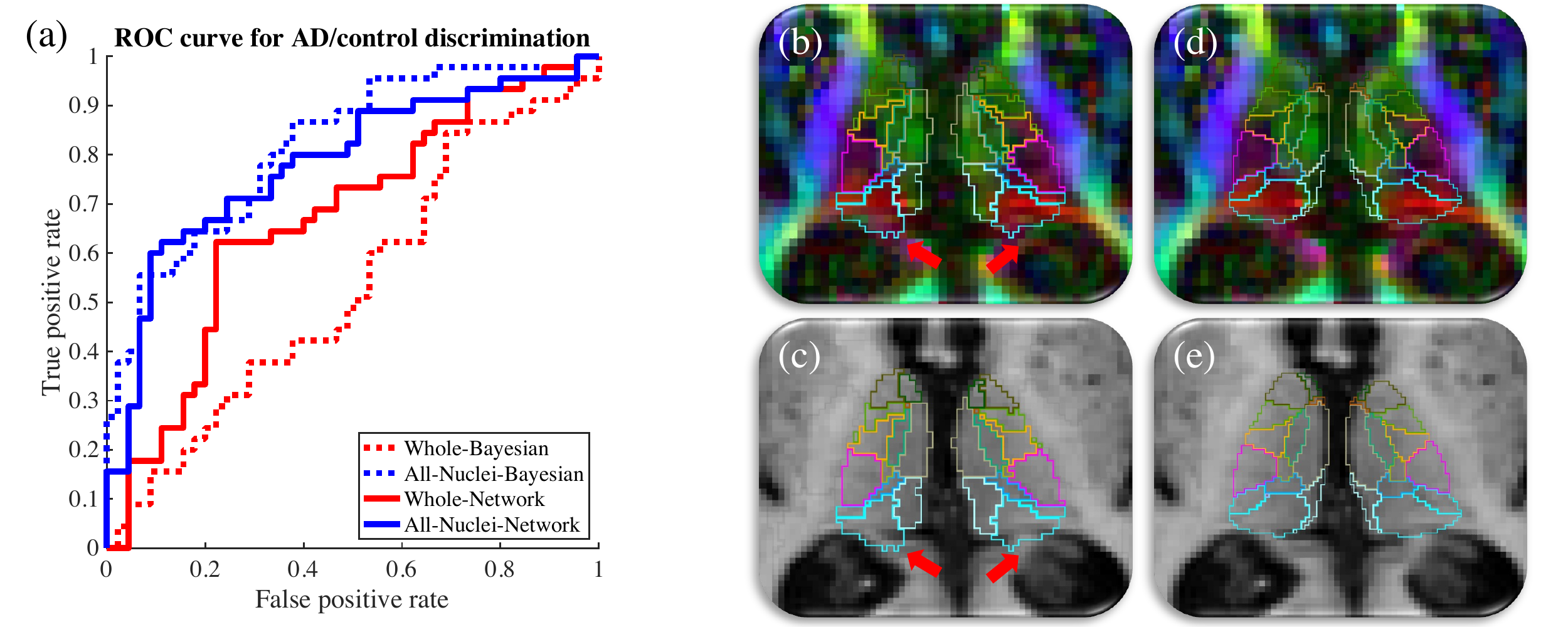}
    \caption{Comparison of Bayesian and CNN segmentation on ADNI. 
    (a)~ROC curves for AD discrimination. (b)~Axial view of RGB encoding of ADNI subject with Bayesian segmentation overlaid. (c)~Corresponding view of T1 scan. (d-e)~Corresponding plane with CNN segmentation.
    Red arrows point at oversegmentation by the Bayesian method.} 
    \label{fig:ADNI}
\end{figure}

\vspace{2mm}
\noindent\textbf{Group study using ADNI:} 
We segmented the ADNI subjects with the best-performing CNN, and computed volumes of the thalamic nuclei normalised by the intracranial volume (estimated with FreeSurfer). We then computed receiver operating characteristic~(\emph{ROC}) curves for AD discrimination using a threshold on: \textit{(i)}~the whole thalamic volume; and \textit{(ii)}~the likelihood ratio given by a linear discriminant analysis (LDA,~\cite{Fisher:1936wq}) on the volumes of the 23 thalamic nuclei (left-right averaged).  The ROC curves are shown in Fig.~\ref{fig:ADNI}(a). The area under the curve (\emph{AUC}) and accuracy at the elbow are shown in Table~\ref{tab:ADNI}.
The LDA of the nuclei volumes from both the CNN and Bayesian methods have similar discriminative power. 
However, there is a sharp increase in the discriminative power of thresholding the whole thalamus volumes from the CNN compared to the Bayesian method.
This indicates the external boundary of our method may be more useful than that provided by the Bayesian method and often corresponds to a reduction of oversegmentation into the pulvinar (Figures~\ref{fig:ADNI}b-e).

\begin{table}[t]
    \centering
    \includegraphics[width=0.9\textwidth]{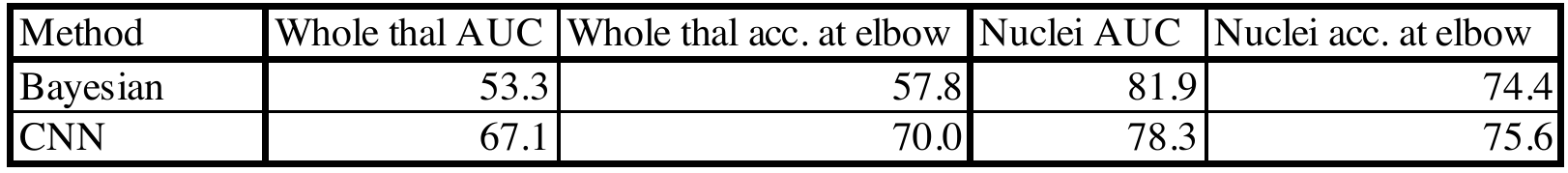}
    \caption{Area under the curve and accuracy at elbow for  AD discrimination. }
    \label{tab:ADNI}
\end{table}

\section{Discussion and conclusion}

We have presented the first method that can segment the thalamic nuclei from T1 and dMRI data obtained with virtually any acquisition, solving the problems posed by PV effects to Bayesian segmentations.
Using Bayesian segmentations generated from multiple diffusion models while applying hybrid domain randomisation and augmentation methods, we remarkably improve upon both the accuracy and reliability of our source segmentations. 
Nuclei volumes resulting from the tool show similar discriminative power to those provided by the Bayesian tool, while improving the utility of whole thalamus measurements and increasing segmentation resolution. 
Crucially, our use of the FA and V1 representation of dMRI data as input means that our tool is compatible with virtually every dMRI dataset. 
Publicly sharing this ready-to-use tool will enable neuroimaging studies of the thalamic nuclei without requiring any expertise in neuroanatomy or machine learning, and without any specialised computational resources.

\section*{Acknowledgments}
This work was primarily funded by Alzheimer’s Research UK (ARUK-IRG2019A003).
PG’s work in this area was supported by NIH NIBIB NAC P41EB015902
AY’s work in this area was supported by NIH grants R01 EB021265 and R56 MH121426.
DCA’s work in this area was supported by EPSRC grant EP/R006032/1 and Wellcome Trust award 221915/Z/20/Z.
The Dementia Research Centre is supported by Alzheimer's Research UK, Alzheimer's Society, Brain Research UK, and The Wolfson Foundation. 
This work was supported by the National Institute for Health Research (NIHR) Queen Square Dementia Biomedical Research Unit and the University College London Hospitals Biomedical Research Centre, the Leonard Wolfson Experimental Neurology Centre (LWENC) Clinical Research Facility, and the UK Dementia Research Institute, which receives its funding from UK DRI Ltd, funded by the UK Medical Research Council, Alzheimer's Society and Alzheimer's Research UK. 
This project has received funding from the European Union’s Horizon 2020 research and innovation program under the Marie Sklodowska-Curie grant agreement No. 765148, as well as from the National Institutes Of Health under project number R01NS112161.
MB is supported by a Fellowship award from the Alzheimer’s Society, UK (AS-JF-19a-004-517). 
MB’s work was also supported by the UK Dementia Research Institute which receives its funding from DRI Ltd, funded by the UK Medical Research Council, Alzheimer’s Society and Alzheimer’s Research UK. 
JDR is supported by the Miriam Marks Brain Research UK Senior Fellowship and has received funding from an MRC Clinician Scientist Fellowship (MR/M008525/1) and the NIHR Rare Disease Translational Research Collaboration (BRC149/NS/MH). 
JEI is supported by the European Research Council (Starting Grant 677697, project BUNGEE-TOOLS) and the NIH (1RF1MH123195-01 and 1R01AG070988-01).

\bibliographystyle{splncs04}
\bibliography{ThalamusCompact}

\end{document}